\documentclass{iau}
\usepackage{graphicx}

\title[Black holes and nuclear star clusters] 
{Black hole and nuclear cluster scaling relations: 
$M_{\rm bh}\propto M_{\rm nc}^{2.7\pm0.7}$}

\author[Alister W.\ Graham]   
{Alister W.\ Graham}

\affiliation{
Centre for Astrophysics and Supercomputing, \\ 
Swinburne University of Technology, \\
Hawthorn, Victoria 3122, Australia. \\ 
email: {\tt AGraham@swin.edu.au}
}

\pubyear{2015}
\volume{312}  
\pagerange{1--5} 
\jname{Star Clusters and Black Holes in Galaxies across Cosmic Time}
\editors{R. Spurzem, A.C. Editor, \& B.D. Editor, eds.}
\begin{document}

\maketitle

\begin{abstract}

There is a growing array of supermassive black hole and nuclear star cluster
scaling relations with their host spheroid, including a bent (black hole
mass)--(host spheroid mass) $M_{\rm bh}$--$M_{\rm sph}$ relation and 
different (massive compact object mass)--(host spheroid velocity dispersion)
$M_{\rm mco}$--$\sigma$ relations for black holes and nuclear star clusters. 
By combining the observed $M_{\rm bh} \propto \sigma^{5.5}$
relation with the observed $M_{\rm nc} \propto \sigma^{1.6–-2.7}$ relation,
we derive the expression $M_{\rm bh} \propto M_{\rm nc}^{2–-3.4}$, which
should hold until the nuclear star clusters are eventually destroyed in the
larger core-S\'ersic spheroids.  This {\it new} mass scaling relation helps 
better quantify the rapid evolutionary growth of massive black holes in
dense star clusters, and the relation is consistently recovered when coupling the
observed $M_{\rm nc} \propto M_{\rm sph}^{0.6–-1.0}$ relation with the recently
observed quadratic relation $M_{\rm bh} \propto M_{\rm sph}^2$ for 
S\'ersic spheroids. 
\end{abstract}

\firstsection 
\section{Introduction}

Over the past two decades there has been wide-spread interest in the scaling
relations connecting supermassive black holes (SMBHs) with their host galaxy,
and in particular with their host bulge.  This has been, in part, due to
observations which suggested that they grow in tandem, with feedback from the black
hole (previously) thought to establish a near constant 0.1--0.2\% mass ratio
with the host spheroid.  Over the last decade there has been a quieter
realisation that the nuclear star clusters (NSCs)\footnote{Nuclear star clusters are
  so-named because of their location at the nuclei of galaxies.} at the
centres of most S\'ersic galaxies also correlate with the properties of their
host spheroid.  This connection continues until the disappearance / destruction
of the clusters in the (massive) core-S\'ersic galaxies with partially
depleted cores (Bekki \& Graham 2010, and references therein).  
Given the coexistence of SMBHs within dense star clusters
(e.g.\ Seth et al.\ 2008; Gonz{\'a}lez Delgado et al.\ 2008; Graham \& Spitler
2009; Graham 2012b; Leigh et al.\ 2012; Neumayer \& Walcher 2012; Scott \&
Graham 2013)
one may wonder if massive black holes might also be intimately
connected with their host star cluster, in addition to 
their host spheroid, or perhaps SMBHs and dense NSCs merely inevitable
co-inhabitants at the bottom of each galaxy's gravitational potential well. 

In this review talk I briefly present the latest scaling relations between both SMBHs
and NSCs with their host spheroid's {\it (i)} velocity dispersion (Section~2)
and {\it ii} stellar mass (Section~3). After then reminding ourselves what
S\'ersic and core-S\'ersic galaxies are (Section~4), these relations are
consistently brought together in a way that eliminates the spheroid and
yields, for the first time, the mass relation between SMBHs and their host
NSCs (Section~5).


\section{The $M_{mco}$-$\sigma$ relations}

Massive black holes and dense nuclear star clusters\footnote{By this term we
  mean to exclude (obvious) nuclear discs, which can be much more extended
  than compact nuclear star clusters (see Balcells et al.\ 2007).} are
collectively referred to here as massive compact objects (mco).  In the
(massive compact object mass)--(host spheroid velocity dispersion)
$M_{mco}$-$\sigma$ diagram, SMBHs and NSCs follow different tracks.

The $M_{\rm bh} \propto \sigma^{X}$ relation has 
a logarithmic slope $X$ of around 5.5$\pm$0.3 
(Graham \& Scott 2013; McConnell \& Ma 2013). 
Galaxies with bars have also been observed to display an apparent 
offset to lower black hole masses in the $M_{\rm bh}$--$\sigma$ diagram 
(Graham 2008; Hu 2008; Graham \& Li 2009; Graham et al.\ 2011). 
As was noted by Hu (2008) and Graham (2008), this may be due to 
under-massive black holes 
in what might be pseudobulges (an idea preferred by Greene et al.\ 2010 and Kormendy \&
Bender 2011), or instead it may be due to the occurrence of higher velocity dispersions. 
Hartmann et al.\ (2013) 
have recently shown that the dynamics associated with bars are indeed
fully capable of explaining the observed offset in the $M_{\rm bh}$--$\sigma$
diagram in terms of elevated velocity
dispersions (see also Brown et al.\ 2013 and Debattista et al.\ 2013), and Graham
\& Scott (2013) have found no offset between barred and unbarred galaxies in
the $M_{\rm bh}$--$L_{\rm sph}$ diagram, disfavouring the pseudobulge idea 
suggested 7 years ago. 

The $M_{\rm nc}$--$\sigma^Y$ relation has a much shallower slope than the
$M_{\rm bh}$--$\sigma^X$ relation. 
Excluding nuclear disks, Graham (2012b) reported a logarithmic slope $Y$ of
$1.57\pm0.24$, 
while Scott \& Graham (2013) reported a value of $2.11\pm0.31$ having 
over-lapping error bars. 
Leigh et al.\ (2012) have however reported a steeper value of $2.73\pm0.29$, 
attributed to their inclusion of nuclear disks which can be an order of
magnitude more massive than the biggest nuclear star clusters.

\section{S\'ersic versus core-S\'ersic galaxies}

S\'ersic galaxies contain spheroids, either bulges or the main elliptical
galaxy itself, whose projected light is well described by S\'ersic's $R^{1/n}$
(1963) model. These S\'ersic spheroids may additionally contain NSCs. 
In contrast, core-S\'ersic galaxies display a partially depleted core, 
not due to dimming by dust and typically less than a few hundred parsec in
radius, relative to the inward extrapolation of their outer S\'ersic profile 
(Graham et al.\ 2003; Trujillo et al.\ 2004).
The S\'ersic versus core-S\'ersic divide built on but differs from the
``core'' versus ``power-law'' galaxy divide (Lauer et al.\ 1995) 
in that ``core'' galaxies 
do not always have a partially depleted core relative to their outer profile
(see Dullo \& Graham 2014, and references therein). 
The core-S\'ersic galaxies are thought to have formed from the dry merger of
S\'ersic (and/or core-S\'ersic) galaxies, wherein the SMBHs sink to the centre 
via the gravitational ejection of stars from the core of the
newly formed galaxy. 

With S\'ersic indices from less than 1 to $\sim$4, S\'ersic spheroids follow a
log-linear luminosity-(central surface brightness) relation ($L-\mu_0$) and a
log-linear $L-n$ relation (e.g.\ Graham \& Guzm\'an 2003, and references
therein).  Due to the non-homology in their light profiles, i.e.\ the fact
that they do not all have the same ($R^{1/4}$, for example) light profiles,
this systematic change in $n$ with luminosity produces a non-linear
luminosity-dependent difference between $\mu_0$ and $\langle \mu \rangle_{\rm
  e}$ (the mean surface brightness within the effective half light radius,
$R_{\rm e}$).  This results in a strongly curved $L-\langle \mu \rangle_{\rm
  e}$ relation.  Given that $L=2\pi\langle I \rangle_{\rm e}R_{\rm e}^2$,
where $\langle I \rangle_{\rm e}$ is the average intensity associated with the
average surface brightness, the $L-R_{\rm e}$ relation is also strongly
curved.  These relations are in fact so curved that the faint ($n \lesssim 2$)
and bright ($n \gtrsim 2$) arms of the relations have, before the consequences
of structural non-homology were known, been mistakenly heralded as evidence
for a dichotomy between faint and bright early-type galaxies (see Graham et
al.\ 2013 for an extended review).

Due to the depleted cores in the core-S\'ersic spheroids (typically $M_B <
-20.5\pm0.75$ mag), they branch off from the $L-\mu_0$ relation toward lower
central surface brightnesses.  Core-S\'ersic and S\'ersic spheroids/galaxies
do however follow the same steep $M_{\rm bh}$--$\sigma$ relation (e.g.\ Graham
\& Scott 2013).

\section{The $M_{mco}$-$M_{\rm sph}$ relations}

Before getting to observations of the $M_{mco}$-$M_{\rm sph}$ relations, 
one can already predict the general behavior in the case of the $M_{bh}$-$M_{\rm
  sph}$ relation.  This is done by noting a transition or bend in
the $L-\sigma$ relation found by Davies et al.\ (1983) such that
low-luminosity early-type galaxies (not pseudobulges) follow the 
relation $L-\sigma^2$ while the high-luminosity galaxies 
($M_B < -20.5$ mag) follow a steeper relation. 
Matkovi\'c \& Guzm\'an (2005) explained the bend 
in terms of S\'ersic versus core-S\'ersic galaxies.  This bend was 
recently shown again as a bend in the  $M_{\rm sph}$--$\sigma$ relation 
by Davies and his collaborators in Cappellari et al.\ (2013). 

Coupled with the log-linear $M_{\rm bh}$--$\sigma$ relation noted in
Section~2, the bent $M_{\rm sph}$--$\sigma$ relation necessitates a bent
$M_{\rm bh}$--$M_{\rm sph}$ relation.  As was pointed out in Graham (2012a),
for things to be consistent there {\it must} be a bent relation rather than
the log-linear relation which had been assumed and claimed for well over a
decade.  This of course introduces a huge change to our understanding of the
physical relation between SMBHs and their host spheroid.

While the bent $M_{\rm bh}$--$M_{\rm sph}$ relation was 
first presented in Graham (2012a) with actual data, 
the black hole masses did not probe very 
far down the mass function, making the discovery somewhat hard to see
(although still statistically significant).  However in
Graham \& Scott (2013), see also Scott, Graham \& Schombert (2013), they were
able to include data down to $M_{\rm bh} \approx 10^6 M_{\odot}$, and in
Graham \& Scott (2014) it reaches down to $10^5 M_{\odot}$ through 
the inclusion of over 100 active galactic nuclei with indirectly measured 
black hole masses.  What the two papers in 2013 confirmed is that S\'ersic 
galaxies follow a near-quadratic $M_{\rm bh}$--$M_{\rm sph}$ relation, i.e.\ a
power-law with a slope close to 2, as predicted in Graham (2012b). 
It is only the core-S\'ersic galaxies,
built from simple additive mergers, which branch off and follow a near-linear 
$M_{\rm bh}$--$M_{\rm sph}$ relation.   As a result, 
fitting a single log-linear relation to samples of S\'ersic and core-S\'ersic
galaxies produces a slope greater than 1 and a relation which is not
optimal for either population. 

The $M_{\rm bh}/M_{\rm sph}$ mass ratio for core-S\'ersic galaxies was found by 
Graham (2012a) to be 0.36\%, double the previously assumed constant value for
all galaxy types, and it was increased to 0.49\% in Graham \& Scott (2013). 
However, due to the quadratic relation for the S\'ersic galaxies, 
their $M_{\rm bh}/M_{\rm sph}$ mass ratio can be far lower. 

The $M_{\rm nc}$--$M_{\rm sph}$ relation was found by Balcells et al.\ (2003)
among the bulges of disk galaxies, and later by Graham \& Guzm\'an (2003)
using a sample of predominantly elliptical galaxies.  The slope of this
relation has since been measured many times, most recently by den Brok et
al.\ (2014) who reports $L_{\rm nc} \propto L_{\rm sph}^{0.57\pm0.05}$
($F814W$), in fair agreement with the value of $0.60\pm0.10$ from Scott \&
Graham (2013) for the $M_{\rm nc}$--$L_{\rm sph}$ ($K$-band) relation for
early-type galaxies.  Previous works have claimed slopes around 0.75 but as
high as 1 when including nuclear disks (e.g.\ Grant et al.\ 2005; Wehner \&
Harris 2006; C\^ot\'e et al.\ 2006; Balcells et al.\ 2007).

\section{The (new) $M_{bh}$-$M_{\rm nc}$ relation}

Coupling $M_{\rm nc} \propto M_{\rm sph}^{0.6–-1.0}$ with 
$M_{\rm bh} \propto M_{\rm sph}^2$ for the S\'ersic spheroids 
gives
$M_{\rm bh} \propto M_{\rm nc}^{2–-3.3}$. 

Coupling 
$M_{\rm bh} \propto \sigma^{5.5}$ with 
$M_{\rm nc} \propto \sigma^{1.6–-2.7}$ 
from Section~2 gives
$M_{\rm bh} \propto M_{\rm nc}^{2.0–-3.4}$. 

Depending on which precise slopes one adopts from the wedded pair of relations
above, one ends up with a different slope for the new relation between black hole
mass and host nuclear star cluster mass.  While the author's past work
would favour a steeper exponent (3.4), encompassing the wider literature suggests
something like $M_{\rm bh} \propto M_{\rm nc}^{2.7\pm0.7}$ given the range of
slopes for the initial relations.  It is hoped that
further observations and analysis, combined with theory, will be able to
refine and explain this steep (non-linear) relation which may not simply be a
consequence of the relations from which it was derived here.

As noted in the Introduction, not all galaxies with SMBHs have
NSCs, and as such a certain degree of common sense is required in application
of this new relation. For instance, core-S\'ersic galaxies do not have a NSC, which was
likely eroded away prior to the formation of their partially-depleted cores. 
Given this, as one approaches the high-mass end of the S\'ersic galaxy 
sequence (from lower masses), it is expected that the NSCs will flay and some S\'ersic galaxies
would no longer house any significant NSC (e.g., NGC~5831, Graham et 
al.\ 2003).  At the low-mass end, and as with the $M_{\rm bh}$--$\sigma$
relation and the $M_{\rm bh}$--$M_{\rm sph}$ relation, the frequency of
massive black holes below $10^5 M_{\odot}$ is not yet known. The occurrence of 
NSCs is however known to tailor off, or at least become harder to identify, in
early-type galaxies fainter than $M_{F814W}=-15$ mag (den Brok et al.\ 2014), which may
then reflect some kind of lower bound to the relation.
Finally, it is remarked that massive BHs in globular clusters may be better matched to
the new $M_{bh}$-$M_{\rm nc}$ relation than the $M_{bh}$-$M_{\rm sph}$ relation.

\section{Conclusions}

SMBHs grow rapidly relative to their stellar nurseries, i.e.\ the nuclear cluster of
stars which still enshroud many.  This is not to say that we know if the SMBHs were born
in these nurseries; although once they become one hundred million
solar mass grown-ups their nursery is gone.  
The growth of the BH relative to
the NSC is extremely rapid: $M_{\rm bh} \propto M_{\rm nc}^{2.7\pm0.7}$, 
with the author favouring higher values for the exponent, especially if new
data steepens the $M_{\rm bh}$--$\sigma$ relation, and if the $M_{bh}$-$M_{\rm
  sph}$ relation {\it is} super-quadratic for the S\'ersic galaxies.

\section{Acknowledgments}

The author thanks the conference organizers for bringing together researchers
of massive black holes and star clusters, and for the opportunity to present
this invited review/update which resulted in the formulation of the steep
(black hole mass)--(nuclear star cluster mass) relation given here.
This research was supported by the Australian Research Council through
funding grant FT110100263.

\end{document}